\documentclass{emulateapj}

\shorttitle{Development of GRMHD Code and Application to GRBs}
\shortauthors{S. Nagataki}

\begin{document}

\title{Development of General Relativistic Magnetohydrodynamic Code
and its Application to Central Engine of Long Gamma-Ray Bursts}

\author{Shigehiro Nagataki\altaffilmark{1}}
\altaffiltext{1}{Yukawa Institute for Theoretical Physics, Kyoto
University, Oiwake-cho Kitashirakawa Sakyo-ku, Kyoto 606-8502, Japan, nagataki@yukawa.kyoto-u.ac.jp}

\begin{abstract}
In order to investigate formation of relativistic jets at the
center of a progenitor of a long gamma-ray burst (GRB), we develop a two-dimensional general
relativistic magnetohydrodynamic (GRMHD) code.
We show the code passes many, well-known test calculations, by which 
the reliability of the code is confirmed. Then 
we perform a numerical simulation of a collapsar using a
realistic progenitor model.
It is shown that a jet is launched from the center of the progenitor.
We also find that the mass accretion rate after the launch of
the jet shows rapid time variability
that resembles to a typical time profile of a GRB.
The structure of the jet is similar to the previous study: a poynting
flux jet is surrounded by a funnel-wall jet.
Even at the final stage of the simulation, bulk Lorentz factor of the
jet is still low, and total energy of the jet is still as small as $10^{48}$ erg. 
However, we find that the energy flux per
unit rest-mass flux is as high as $10^2$ at the bottom of the jet. 
Thus we conclude that the bulk Lorentz factor of the jet can be
potentially high when it propagates outward.
It is shown that the
outgoing poynting flux exists at the horizon around the polar region,
which proves that the Blandford-Znajek mechanism is working. 
However, we conclude that the jet is launched mainly by
the magnetic field amplified by the gravitational collapse and
differential rotation around the black hole, rather than the
Blandford-Znajek mechanism. 
\end{abstract}

\section{Introduction}

Gamma-Ray Bursts (GRBs; in this study, we
consider only long GRBs, so we refer to long GRBs as GRBs hereafter)
have been mysterious phenomena since their discovery
in 1969~\cite{klebesadel73}. Last decade, observational evidence for
supernovae (SNe) and GRBs  
association has been reported (e.g. Woosley and Bloom
2006, and references therein).

Some of the SNe that associate with GRBs were very energetic and
blight. The estimated explosion energy was of the order of
$10^{52}$ ergs, and produced nickel mass was $\sim 0.5M_{\odot}$.
Thus they are categorized as a new type of SNe (sometimes
called as hypernovae). 
The largeness of the explosion energy is very important, because
it can not be explained by the standard core-collapse SN scenario,
and other mechanism should be working at the center of the progenitors.

The promising scenarios are the collapsar
scenario~\cite{woosley93} and the magnetar scenario~\citep{usov92}. 
In the collapsar scenario, a rapidly rotating black hole (BH) is
formed at the center, while a rapidly rotating neutron star with
strong magnetic fields ($\sim 10^{15}$G) is formed in the magnetar
scenario. Many numerical simulations have been done for the collapsar
scenario~\citep{macfadyen99,proga03,proga03b,mizuno04a,mizuno04b,proga05,fujimoto06,shibata06,nagataki07,sekiguchi07,suwa07,barkov08}
and the magnetar
scenario~\citep{takiwaki04,komissarov07,burrows07,bucciantini08,dessart08,takiwaki08,bucciantini09}.
In this study, we investigate the collapsar scenario.

In the collapsar scenario, a BH is formed as a result of
gravitational collapse. Also, rotation of the progenitor plays an
essential role. Due to the rotation, an accretion disk is formed
around the equatorial plane. On the other hand, the matter around the
rotation axis falls into the BH almost freely. It is pointed out that
the jet-induced explosion along the rotation axis may occur
due to the heating through pair annihilation of neutrinos and anti-neutrinos 
that are emitted from the accretion
disk~\citep{woosley93,macfadyen99,fryer00}. Effect of extraction
of rotation energy from the accretion disk by magnetic 
field lines that leave the disk surface (Blandford-Payne
effect~\cite{blandford82}) is also investigated by several
authors~\citep{proga03,proga03b,mizuno04a,mizuno04b,proga05,fujimoto06,nagataki07,suwa07}. 
Recently, the effect of extraction
of rotation energy from the BH through outgoing poynting
flux (Blandford-Znajek effect~\cite{blandford77}) is
investigated~\cite{barkov08}. 
In order to investigate the collapsar scenario completely, a high-quality
numerical code including effects of a lot of microphysics
(neutrino physics, nuclear physics, and equation of state for dense
matter) and macrophysics (magneto-hydrodynamics, general relativity)
has to be developed. Although many numerical studies have been
reported, such a numerical code has not been developed yet. Thus we have
to develop our numerical code step by step.

In this study, we investigate the dynamics of collapsars taking into
account the general relativistic effects. 
Extraction of rotation energy from a rotating BH is one of them.
Also, even when the rotation energy is extracted from the accretion
disk, the properties of the accretion disk should depend on the
properties of the BH: if the BH is rotating, the inner
region of the accretion disk should be enforced to co-rotates
with the BH. We investigate how a jet is launched
at the center of a progenitor, and how the property of the jet is.
Effects of rotation of the BH on the
formation of GRB jet have not been investigated so much. Barkov and
Komissarov (2008) is a pioneering study. However, only one case is
investigated in their study, and the initial progenitor model they
used is a simplified one-dimensional model without rotation and magnetic
fields~\cite{bethe90}. Since there should be many initial
conditions of progenitors (progenitor mass, metallicity, angular
momentum, magnetic fields), it should be important to investigate the
general relativistic effects using a different initial condition from the previous study.
In this study, we use a realistic initial condition for the progenitor
model that is developed by Woosley and Heger (2006), in which
rotation and magnetic fields are taken into account.

When we investigate the general relativistic effects, one has to develop
a General Relativistic Magneto-Hydro Dynamic (GRMHD) code. 
So far, there are many studies on GRMHD code for fixed background
space times using high-order conservative schemes based on either
approximate or full wave-decomposition Riemann
solvers~\citep{gammie03,komissarov05,anninos05,anton06,delzanna07,tchekhovskoy07} 
or non-conservative schemes~\citep{devilliers03,anninos05}.
Since the accreted mass onto the BH is still less than the
initial BH mass in this study, we take the GRMHD code for the
fixed background. Especially, we develop our code using the
conservative scheme of Gammie et al. (2003) with the method of Noble et
al. (2006) for transforming conserved variables to primitive 
variables.

The plan of the paper is as follows. In section~\ref{GRMHD1}, we
present the formulation of the GRMHD code. In section~\ref{test1}, we
show results of many, well-known test calculations to confirm the
reliability of the code. After we show the reliability, we present
results of numerical simulations of collapsars in section~\ref{GRB1}.
Summary and discussion are presented in section~\ref{summary}.

\section{Development of GRMHD Code}\label{GRMHD1}

We have developed a two-dimensional
GRMHD code following Gammie et al. (2003) and Noble et al. (2006).
We have adopted a conservative, shock-capturing scheme with Harten,
Lax, and van Leer (HLL) flux term~\cite{harten83} with
flux-interpolated constrained transport technique~\cite{toth00}.
We use a third-order Total Variation Diminishing (TVD)
Runge-Kutta method for evolution in time,
while monotonized central slope-limited linear interpolation method is
used for second-order accuracy in space~\cite{van77}.
2D scheme (2-dimensional Newton-Raphson method) is usually adopted for 
transforming conserved variables to primitive 
variables~\cite{noble06}.

When we perform simulations of GRMHD,
Modified Kerr-Schild coordinate is basically adopted with mass of the
BH ($M$) fixed where the Kerr-Schild radius $r$ is replaced by the
logarithmic radial coordinate $x_1= \ln r$. 
When we show the result, the coordinates are sometimes transfered from
Modified Kerr-Schild coordinate to Kerr-Schild one for convenience.
In the following, we use $G=M=c=1$ unit.
$G$ is the
gravitational constant, $c$ is the speed of light, and $M$ is the
gravitational mass of the BH at the center.
Throughout this paper we follow the standard notation~\cite{MTW70}.

\subsection{Formalism}\label{GRMHD2}

Number of variables that appear in the equations of GRMHD
is 13: rest-mass density ($\rho$), internal energy density ($u$),
pressure ($p$), four-velocity of fluid ($u^{\mu}$), and Faraday
tensor ($F^{\mu \nu}$). Note that Faraday tensor has only 6 independent
components due to the relation $F^{\mu \nu} = - F^{\nu \mu}$.
We can reduce the number of independent variables to 8 using 
the MHD condition ($u_{\mu} F^{\mu \nu} =0$), equation of state
($p = (\gamma -1)u$: $\gamma$-law gas is assumed),
and the unit length of the four velocity
($u_{\mu}u^{\mu} = -1$). Note that the number of independent equations
of the MHD condition is 3. We choose ($\rho$,$u$,$u^{i}$,$B^i$)
as the 8 independent variables where $u^{i}$ is the space component
of the four velocity. $B^i$ can be written as $\Re^i$/$\alpha$ where
$\alpha$ is the lapse function ($\alpha = \sqrt{-1/g^{tt}}$) and
$\Re^i$ is the magnetic field measured by the Fiducial observer (FIDO)
whose four velocity is $n_\mu = (- \alpha,0,0,0)$. We call these
independent variables as the primitive variables. Below, we introduce
the conserved variables. Of course, number of the conserved
variables is also 8. Thus we require 8 basic equations to follow the
time evolution of the system.

The basic equations of GRMHD represent the rest-mass conservation,
the energy-momentum conservation, and space component of the induction
equation that determines the time evolution of the magnetic fields. 
These are:
\begin{equation}
\partial_t (\sqrt{-g} \rho u^t) = - \partial_i (\sqrt{-g} \rho u^i)
\label{eq1}
\end{equation}
\begin{equation}
\partial_t (\sqrt{-g} T^t_\nu) = - \partial_i (\sqrt{-g} T^i_\nu) +
\sqrt{-g} T^\kappa_\lambda \Gamma^\lambda_{\nu \kappa}
\label{eq2}
\end{equation}
\begin{equation}
\partial_t (\sqrt{-g} B^i) = - \partial_j \left[ \sqrt{-g}( b^iu^j -
b^ju^i) \right],
\label{eq3}
\end{equation}
where $T^{\mu \nu}$ is the stress energy tensor that is composed of
the sum of the matter part ($T^{\mu \nu}_{\rm Matter} = (\rho + u +
p)u^\mu u^\nu + p g^{\mu \nu}$) and electromagnetic part ($T^{\mu
\nu}_{\rm EM} = F^{\mu \alpha} F^{\nu}_{\alpha} - g^{\mu \nu}
F_{\alpha \beta} F^{\alpha \beta}/4$). The factor of $\sqrt{4 \pi}$ is
absorbed into the definition of the Faraday tensor~\cite{gammie03}.
$b^{\mu}$ is introduced so that Eq.(\ref{eq3}) looks
simple, and it is defined as $b^{\mu} = \epsilon^{\mu \nu \kappa
\lambda} u_\nu F_{\lambda \kappa}$ where $\epsilon^{\mu \nu \kappa
\lambda} = (-1/\sqrt{-g}) \left[   \mu \nu \lambda \kappa
\right]$. $\left[   \mu \nu \lambda \kappa \right]$ is the
completely antisymmetric symbol. In the fluid-rest frame, $b^{\mu}$
becomes (0,$B^i$).

In this study, we adopt the conservative scheme for integration
of the GRMHD equations. In this case, the left terms of
Eq.(\ref{eq1})-(\ref{eq3}) are considered to be fundamental variables
and called as the conserved variables. The right terms of
Eq.(\ref{eq1})-(\ref{eq3}) are flux terms with a source term (the
second right term of Eq.(\ref{eq2})).  

Since we have to estimate pressure of the fluid, we have to estimate
the primitive variables from the conserved variables at each time step.
The problem is that the primitive variables can not be expressed
analytically by the conserved variables. Thus we have to use the
Newton-Raphson method to obtain the primitive variables from the
conserved ones~\cite{noble06}. 

Basically, we adopt the 2D scheme introduced by 
Noble et al. (2006) to calculate the primitive variables.
However, it sometimes happens that the 2D scheme fails to
converge well, and the primitive variables can not be obtained
precisely. In such a case, we first adopt the 1$\rm D_W$ scheme
introduced by Noble et al. (2006) and see whether the 1$\rm D_W$
scheme converges. If it converges well, we adopt the primitive
variables obtained by the 1$\rm D_W$ scheme for the next time step.
Otherwise, we adopt the second choice explained in the following subsection.

\subsection{Supplemental Method to Calculate Primitive
Variables}\label{GRMHD3} 

Following Noble et al. (2006), we introduce convenient variables
$v^2$, $W$, $Q^{\mu}$, and $\bar{Q}^\mu$. These variables, of course,
depend on the primitive variables. The definition of these variables
are: $v^2 = v_i v^i$, $W=\omega \Gamma^2$, $Q^\mu = \alpha T^{t \mu}$,
and $\bar{Q}^\mu$ = $j^\mu_\lambda Q^\lambda$, where $v^i$ is the fluid
velocity relative to FIDO, $\omega = \rho + u + p$, $\Gamma =
1/\sqrt{1-v^2}$, and $j_{\mu \lambda} = g_{\mu \lambda} + n_\mu
n_\lambda $. 
It is apparent that $Q^{\mu}$ and $\bar{Q}^\mu$ can be written
analytically by the 
conserved variables. On the other hand, $v^2$ and $W$ can not be
expressed analytically by the conserved variables. 
Thus, we have to solve $v^2$ and $W$ numerically in order to
determine the proper, corresponding primitive variables. 

Here we show that an upper limit and a lower limit for $W$ can be
obtained before searching for a solution of $W$ and $v^2$ numerically.
Thanks to this fact, all
we have to do is to seek the solution with the condition $W_{\rm min}
\leqq W \leqq W_{\rm max}$. From Eq.(28) and Eq.(29) in Noble et
al. (2006), $W$ and $v^2$ satisfy the following equations:
\begin{equation}
v^2_{\rm eq28} = \frac{ \bar{Q}^2W^2 + (Q_\mu \Re^\mu)^2(\Re^2 + 2W)   }{(\Re^2 +
W)^2 W^2}  
\label{noble28}
\end{equation}
\begin{equation}
v^2_{\rm eq29} = \frac{2}{\Re^2} \left[  \frac{(Q_\mu \Re^\mu)^2}{2
W^2} - W + p - (Q_\mu n^\mu)  \right] -1. 
\label{noble29}
\end{equation}
From these equations and the relation $0 \leqq v^2 < 1$, 
$v^2$ and $W$ satisfy the following relations:
\begin{eqnarray}
\nonumber
f(W) &=& W^4 + 2 \Re^2W^3 + (\Re^4 - \bar{Q}^2)W^2 - 2(Q_\mu \Re^\mu)^2
W \\ 
&-& \Re^2 (Q_\mu \Re^\mu)^2 \geqq 0
\label{eq4}
\end{eqnarray}
\begin{equation}
g(W) = W^3 + \{ \frac{1}{2} \Re^2 + (Q_\mu n^\mu) -p\} W^2 -
\frac{1}{2}(Q_\mu \Re^\mu)^2 \leqq 0 
\label{eq5}
\end{equation}
\begin{equation}
h(W) = W^3 + \{ \Re^2 + (Q_\mu n^\mu) -p\} W^2 -
\frac{1}{2}(Q_\mu \Re^\mu)^2 \geqq 0. 
\label{eq6}
\end{equation}
Since $f(0) \leqq 0$, $f^{'}(0) \leqq 0$, and at least one of the
solution for $f^{''}(W) = 0$ is less than 0, there is only one
positive solution $W_a$ that satisfies $f(W_a) = 0$. Thus, from
Eq.(\ref{eq4}), $W$ has to be greater than $W_a$.

We can understand the behavior of $g(W)$ from its first
derivative for $W$:
\begin{equation}
g^{'}(W) = W \left[ 3W + 2 \{ \frac{1}{2} \Re^2 + (Q_\mu n^\mu) -p\}
\right].  
\label{eq7}
\end{equation}
It is apparent that $W=0$ is a solution for $g^{'}(W)=0$.
As for the other solution(s), it is not so obvious because the
pressure $p$ depends on $W$ and $v^2$. However, it will be natural to
consider that the monotonic relation holds between $W$ and $p$. It
means that the pressure rises when $W$ becomes larger. If this
assumption is adopted, as long as $g^{'}(W)=0$ has another solution, 
it is a positive one $W = W_\alpha \geqq 0$. This is because when
$W=0$, $p$ should be also 0 and $\left[ 3W + 2 \{ 1/2 \Re^2 +
(Q_\mu n^\mu) -p\}\right]$ is a positive value. Thus, $g(W)=0$ has
only one positive solution $W_b$. This holds even if
$g^{'}(W)=0$ has only one solution at $W=0$. Also, same conclusion can be
derived for $h(W)$: there is only one positive solution $W_c$ that
satisfies $h(W_c)=0$.

Since $h(W) \geqq g(W)$, the relation $W_c \leqq W_b$ holds. In
conclusion, $W$ has to be in the range $W_{\rm min}= {\rm
Max}(W_a,W_c) \leqq W \leqq W_b= W_{\rm max}$. Thus all we have to do
is to find a solution of $W$ that satisfies $v^2_{\rm eq28} = v^2_{\rm
eq29}$ in this range. This procedure is more expensive than the 2D scheme
and the 1$\rm D_W$ scheme, but the solution for $W$ and $v^2$ is more likely to
be found because the range for the solution of $W$ is determined apriori.
Thus we use this method as a supplementary one to obtain the primitive
variables. 

\begin{figure}
\epsscale{1.20}
\plotone{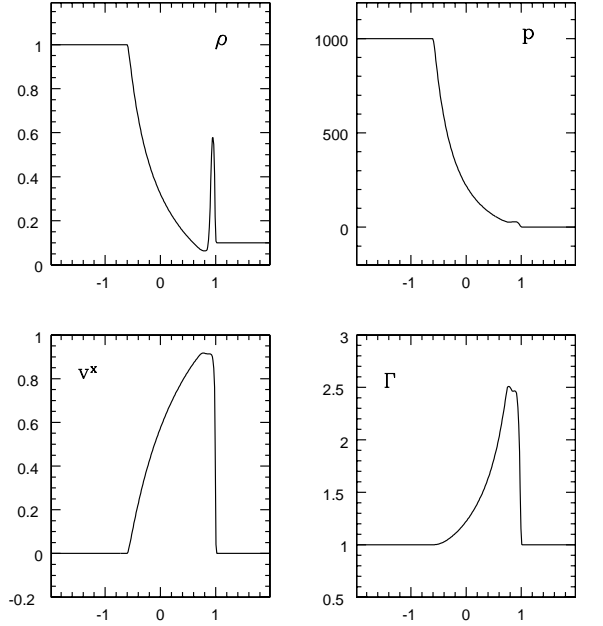}
\caption{Simulation of 1D shock tube test (Komissarov 1999). The state
at $t=1.0$ is shown in the figure. Number of grid points is 600. The
calculation region is set to be $-2 \le x \le 2$. The
upper left panel shows density, the upper right panel shows pressure,
the lower left panel shows the velocity in the x-direction, and the
lower right panel shows the bulk Lorentz factor. 
\label{fig1}}
\end{figure}

\begin{figure}
\epsscale{1.20}
\plotone{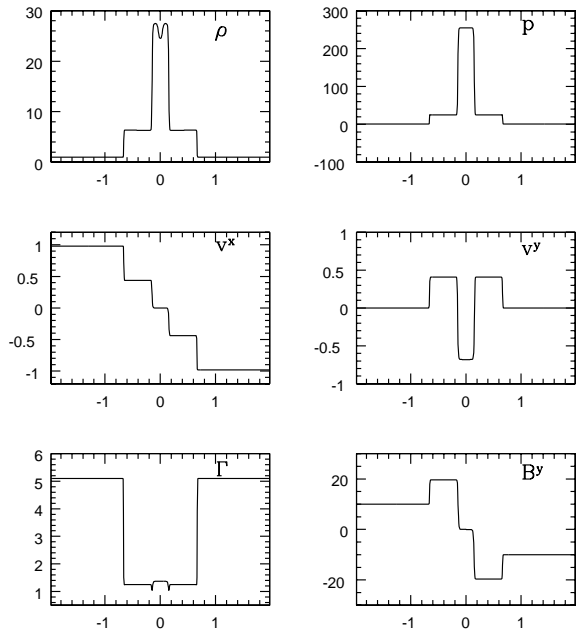}
\caption{Simulation of 1D collision test (Komissarov 1999). The state
at $t=1.2$ is shown in the figure. Number of grid points is 600. The
calculation region is set to be $-2 \le x \le 2$. The left panels show
density, velocity in the x-direction, and bulk Lorentz factor (from
top to bottom), while right panels show pressure, velocity in
x-direction, and y-component of magnetic field (from top to bottom).  
\label{fig2}}
\end{figure}

\section{Test Calculations}\label{test1}
Using the GRMHD code that is developed in this study, we check
whether it can pass many, well-known test calculations. 
The first three tests are special relativistic hydrodynamic (SRHD) or
special relativistic magnetohydrodynamic (SRMHD) calculations, while
the rest of three tests are GRMHD ones.

\subsection{Shock Tube Problems}\label{test2}
1D shock tube tests are the most basic test problems for
SRHD/SRMHD. We have carried out a number of the test simulations
introduced in Komissarov (1999) and Balsara (2001). Here we describe
only two of them. One is the shock tube test1 and the other is the
collision test~\citep{komissarov99,mizuno06}. 
The initial left and right states are summarized in Table.\ref{tab1}.
Number of grid points is 600 for both simulations. 
The results are shown in Fig.\ref{fig1} and Fig.\ref{fig2}, which show
that the test
calculations are well solved as in the previous studies.

\subsection{Double Shock Problems}\label{test3}
\begin{figure}
\epsscale{1.20}
\plotone{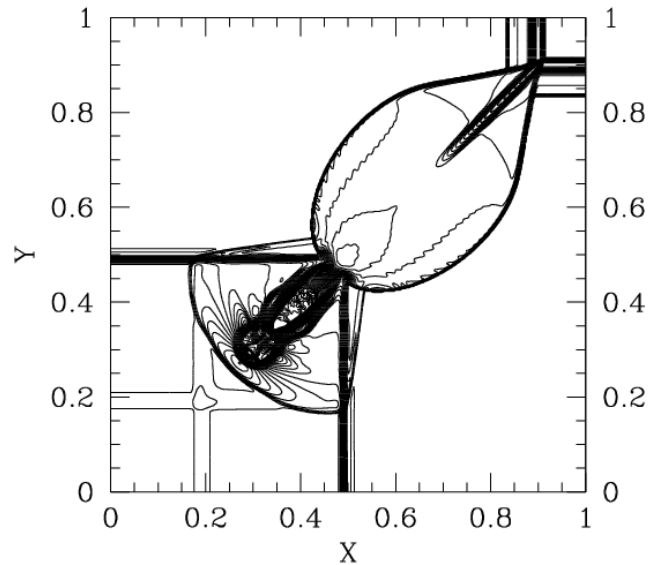}
\caption{Simulation of 2D shock tube problem. Density contour at $t=0.4$ is
shown in the figure. Numbers of grid points are 400$\times$400. The
calculation region is set to be $0 \le x \le 1$ and $0 \le y \le 1$. 
\label{fig3}}
\end{figure}

Here 2D shock tube problem is done to confirm whether the shock dynamics in the
multidimensional flow can be solved safely. This problem includes the
interactions of shocks, rarefactions, contact discontinuities. 
Initially a square computational domain is prepared in x-y plane and
divided into four quarter boxes. Initial condition in each box is
summarized in Table.\ref{tab2}. This condition is same with previous
study~\citep{delzanna02,zhang06,mizuta06}. We use 400$\times$400
uniform grid points in a square computational box. Boundary
condition is open ones.
Density contour at the final stage of the simulation is shown in
Fig.\ref{fig3}, which shows that our code
reproduces the previous studies very well.

\subsection{Cylindrical Explosion Test}\label{test5}
\begin{figure}
\epsscale{1.20}
\plotone{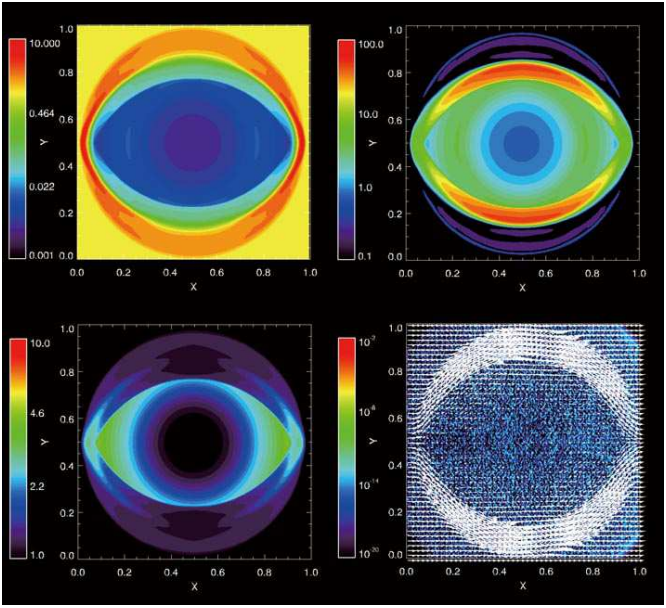}
\caption{A RMHD 2D cylindrical explosion test with a pressure jump as
high as $10^5$. The state at $t=0.4$ is shown in the figure. Numbers of
grid points are 250$\times$250. The
calculation region is set to be $0 \le x \le 1$ and $0 \le y \le
1$. The upper left panel shows the density contour in logarithmic
scale. The upper right panel shows the pressure contour in logarithmic
scale. The lower left panel shows contour of the
bulk Lorentz factor. The lower right
panel shows divergence of the magnetic fields in logarithmic
scale with magnetic field lines.  
\label{fig4}}
\end{figure}

Here we go to a SRMHD test. A famous,
cylindrical blast explosion test is done~\citep{delzanna03,leismann05}. 
We use the $[0,1] \times [0,1]$ Cartesian grid with a resolution of
$N_x=N_y=250$ grid points. We define an initially static background
with $\rho = 1.0, p = 0.01$, and $B_x=4.0$. The relativistic flow
comes out by setting a much higher pressure, $p=10^3$ within a circle
of radius $r=0.08$ placed at the center of the domain. $\gamma$ for
the equation of state is set to be 4/3. Final time is set to be 0.4. 
The result is shown in Fig.\ref{fig4}. The upper left panel shows the
density contour in logarithmic
scale. The upper right panel shows the pressure contour in logarithmic
scale. The lower left panel shows contour of the
bulk Lorentz factor. The lower right
panel shows the divergence of the magnetic fields in logarithmic
scale with magnetic field lines.
These results are consistent with the previous studies. Especially,
the divergence of the magnetic fields is kept as small as $10^{-14}$.

\subsection{Gammie's Flow}\label{test4}
\begin{figure}
\epsscale{1.20}
\plotone{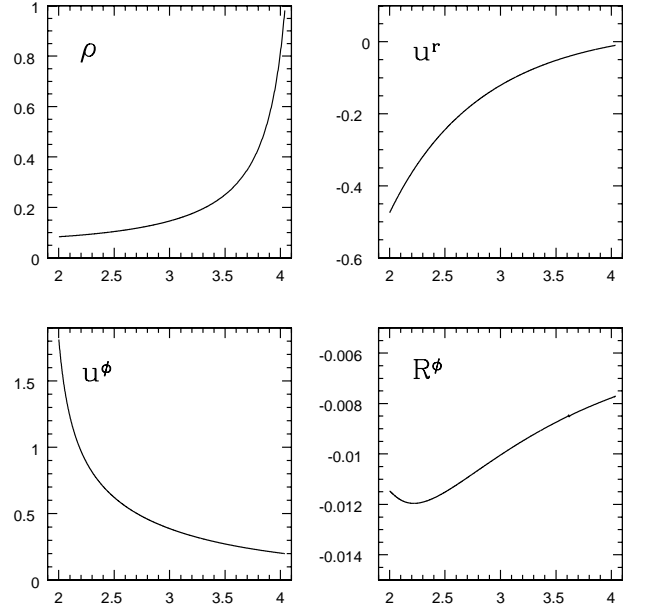}
\caption{Gammie's equatorial inflow solution in the Kerr metric with
$a=0.5$ and magnetization parameter $F_{\theta \phi}=0.5$. Number of
grid point is 1024. The state at $t=1.5$ is shown in the figure. The
panels show density, radial component of the 
4-velocity, the $\phi$ component of the 4-velocity, and $\Re^{\phi}$
at the final stage of the simulation. Boyer-Lindquist coordinate is
used for the simulation. 
\label{fig5}}
\end{figure}

\begin{figure}
\epsscale{1.20}
\plotone{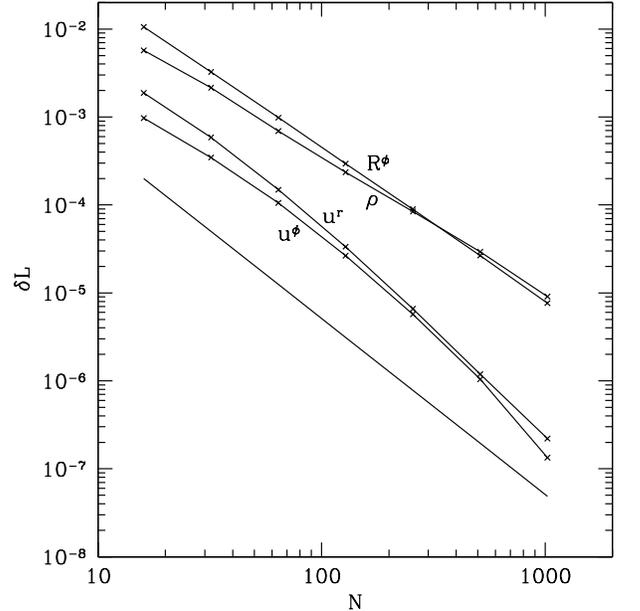}
\caption{Convergence results for the Gammie's equatorial inflow
solution in the Kerr metric with $a=0.5$ and magnetization parameter
$F_{\theta \phi}=0.5$. Norms of the error for $\rho$, $u^r$,
$u^{\phi}$, and $\Re^{\phi}$ at the final stage of the simulation are
shown in the figure. The straight line represents the slope expected for
second-order convergence. The definition of the norm of the error is
written in the text.  
\label{fig6}}
\end{figure}

Next we consider a GRMHD test.
A steady, magnetized inflow solution on the equatorial plane
around a Kerr BH is considered~\citep{takahashi90,gammie99}.
Initially, the steady inflow solution for the Kerr parameter $a=0.5$ and
the magnetization parameter $F_{\theta \phi}=0.5$ is set,
and time evolution of the system is followed by the GRMHD code.
In this calculation, Boyer-Lindquist coordinate is used. The
calculation region is set to be [2.0 $\leqq r \leqq$ 4.04] and [$0.5 -
10^{-3} \leqq \theta/\pi \leqq 0.5 + 10^{-3}$]. The model is run for
$t=1.5$. The physical values at boundaries are fixed throughout the
simulation. Results are shown in Fig.\ref{fig5}: density, radial
component of the
4-velocity, the $\phi$ component of the 4-velocity, and $\Re^{\phi}$
at the final stage of the simulation. When the initial state
is written in the same figure, we can see that the final state
coincides with the initial state. To show it more quantitatively,
we introduce the norms of the errors for these values as a function of the
number ($N$) of grid points in the radial coordinate. The definition
of the norm of the error is $\Sigma_{i=1}^{i=N} \left| a({\rm final})
- a({\rm initial}) 
\right| / \Sigma_{i=1}^{i=N} \left| a({\rm initial}) \right|$. 
In Fig.\ref{fig6}, the norms of errors are shown. We can see that these values
converge roughly proportional to $N^{-2}$, as expected.

\subsection{Blandford-Znajek Monopole Solution}\label{test6}
Further we continue to test the GRMHD code. We consider the
Blandford-Znajek monopole solution~\cite{blandford77}. This analytic
solution
has been investigated numerically by previous
studies~\citep{komissarov04b,mckinney04,tanabe08}.

The computational domain is axisymmetric, with a grid that extends
from $r_{\rm in} = 0.98 r_+$ to $r_{\rm out} = 230$ and from $\theta =
0$ to $\theta = \pi$ where $r_+ = 1 + \sqrt{1-a^2}$ is the outer event horizon.
The numerical resolution is 300 $\times$ 300. 
As an initial condition, we put the 0th order terms of the monopole
solution around the BH~\cite{komissarov04b}. That is,
$\Re^{\mu} = - n_{\nu}
^{*}F^{\mu \nu} = (0,\alpha \sin \theta / \sqrt{-g},0,0)$ in the
Kerr-Schild coordinate where $^{*}F^{\mu \nu}$ and $g$
are the dual field tensor and determinant of the Kerr-Schild metric.
The plasma velocity relative to the FIDO is set to
zero initially, and its pressure and density are set to small value
($P = \rho = \Re^2/100$) so that the system becomes Force-Free like. 
Also, to keep the magnetization reasonably low, when the critical
condition $0.01B^2 \ge \Gamma^2 \rho + (\gamma \Gamma^2 - (\gamma-1)) u$
is satisfied, density and internal energy are increased by the same
factor so that the critical condition holds~\cite{komissarov04b}.
$\gamma$ is set to be 4/3.
We have performed numerical simulations with
the Kerr parameters 0, 0.01, 0.1, 0.2, 0.3, 0.4, 0.5, 0.6, 0.7, 0.8,
0.9, 0.95, 0.99, and 0.995 until time $t$ = 200.   

\begin{figure}
\epsscale{1.20}
\plotone{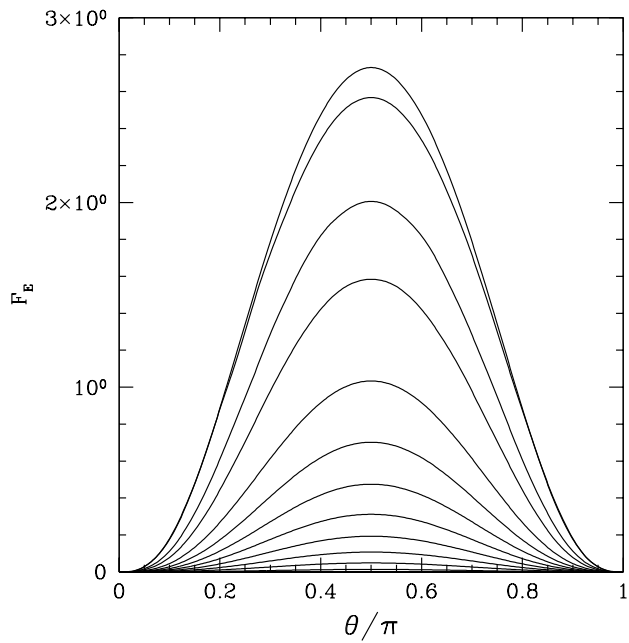}
\caption{Outgoing poynting fluxes as a function of the zenith angle for
the Blandford-Znajek monopole solution with Kerr parameter $a=0.01,
0.1, 0.15, 0.2, 0.3, 0.4, 0.5, 0.6, 0.7, 0.8, 0.9$, and $0.95$. The
fluxes are measured at $r=20$ and $t=200$.
Numbers of grid point are 300$\times$300.
\label{fig7}}
\end{figure}


The total energy flux, which is the integrated outgoing poynting flux
over the zenith angle, can be written as
\begin{eqnarray}
\nonumber
\dot{E} &=& 2 \pi \int_0^{\pi} d \theta \sqrt{-g} (-T^r_t) 
        = 2 \pi \int_0^1     d x_2 \left(  \frac{d \theta}{dx_2}
        \right) \sqrt{-g} (-T^r_t) \\ 
        &=& 2 \pi \int_0^1     d x_2 F_E,
\label{BZeq1}
\end{eqnarray}
where $x_2 = \theta / \pi$ is introduced as a convenient
variable~\cite{gammie03}.

In Fig.\ref{fig7}, the outgoing poynting fluxes ($F_E$) as a function of
zenith angle are shown. The
fluxes are measured at $r=20$ at the final stage of the simulations.
We would like to note that the outgoing poynting flux hardly depends
on the radius where it is evaluated. This means that the conservation of
the outgoing poynting flux is confirmed numerically.

\begin{figure}
\epsscale{1.20}
\plotone{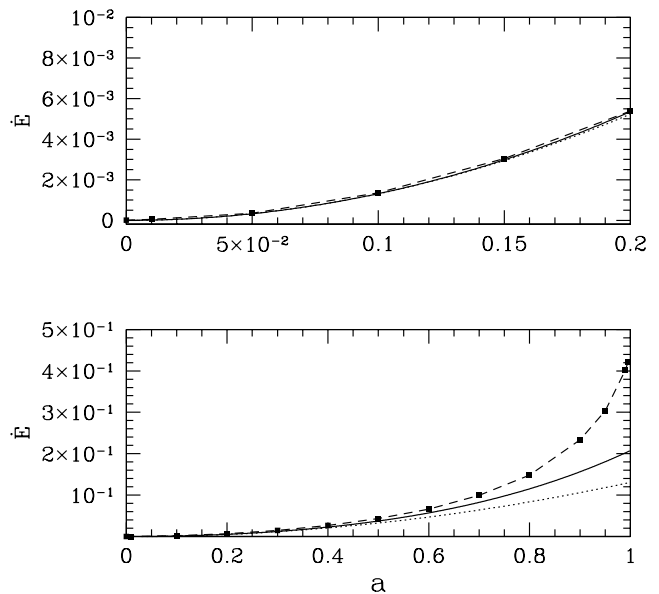}
\caption{Upper panel (a): Comparison of the derived, conserved, total energy flux.
Dashed line with rectangular points is numerical result for small Kerr
parameter ($0 \le a \le 0.2$), dotted line shows the second-order analytical
solution, and solid line represents the forth-order analytical
solution. Lower panel (b): Same with upper panel, but for wide range of the Kerr
parameters $(0 \le a \le 1)$. Simulations are done for the Kerr
parameters 0, 0.01, 0.1, 0.2, 0.3, 0.4, 0.5, 0.6, 0.7, 0.8,
0.9, 0.95, 0.99, and 0.995 until $t=200$.
\label{fig8}}
\end{figure}

In Fig.\ref{fig8}(a), we plot the total energy flux ($\dot{E}$) at the
final stage 
for small Kerr parameters ($0 \le a \le 0.2$) by rectangular points. 
Dashed line is just
the interpolation of the calculated values. For comparison, the
second-order analytical solution is shown by dotted line
and the forth-order analytical solution is shown by solid line.
From this comparison, we can see that all of them coincide with each
other. Thus the results of the numerical simulations by the GRMHD code
are confirmed by analytical solutions.

The situation becomes different for large Kerr parameters. 
In Fig.\ref{fig8}(b), we plot the same values with Fig.~\ref{fig8}(a), but
for wide range of the Kerr parameters ($0 \le a \le 1$). We can see
clearly the difference among three cases. This is because the
analytical solution is obtained by the perturbation method in Kerr
parameter, and it is applicable only for small Kerr parameters.
Of course, there is no such limitation for the numerical simulations.
Thus the total energy flux obtained by the numerical simulation is
more reliable than the analytical estimation (see Tanabe and Nagataki
(2008) for detailed discussion).


\subsection{Fishbone and Moncrief's Test}\label{test7}
Here we present a final test of the GRMHD code. A steady and
stationary torus~\citep{fishbone76,abramowicz78} around a Kerr BH that is supported by both centrifugal force and pressure is
solved numerically. Of course, it should be solved as a steady and
stationary state. 

\begin{figure}
\epsscale{1.20}
\plotone{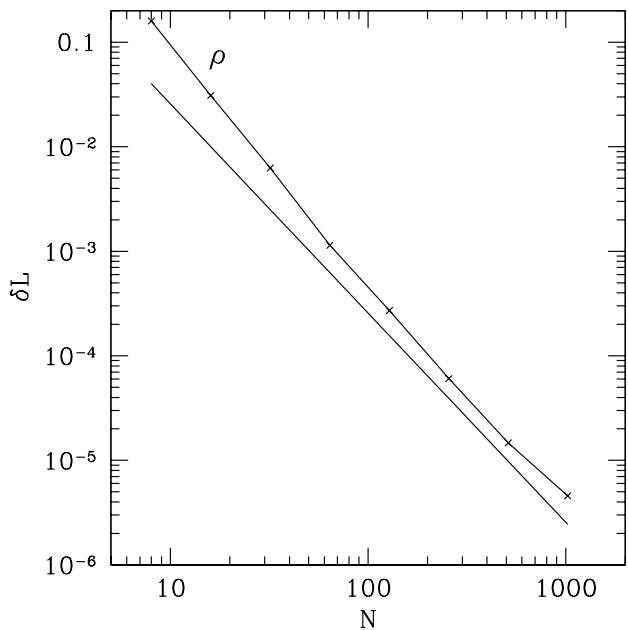}
\caption{Convergence results for the Fishbone-Moncrief problem for a
Kerr BH with $a=0.9$. The straight line represents the slope expected for
second-order convergence. Kerr-Schild coordinate is used for the simulation. 
\label{fig9}}
\end{figure}

We have integrated a Fishbone-Moncrief solution around a Kerr BH with $a=0.9$. We set $u^t u_\phi = 4.45$ and $R_{\rm in}=6.0$. 
The grid extends radially from $r_{\rm in} = 1.40$ to $r_{\rm
out}=100$. The same floors with Gammie et al. (2003) are used for $\rho$
and $u$. The numerical resolution is $N \times N$ and the solution is
integrated for $t=10$. The resulting norm of the error, which
converges roughly proportional to $N^{-2}$, is shown
in Fig.\ref{fig9}.

Next we follow the time evolution of the Fishbone-Moncrief solution
with magnetic fields.
The vector potential, $A_\phi \propto {\rm max} (\rho/\rho_{\rm
max} - 0.2,0)$ where $\rho_{\rm max}$ is the peak density in the torus,
is introduced~\cite{gammie03}. The field is normalized so that the
minimum value of $p_{\rm gas}/p_{\rm mag}$ becomes $10^2$. The time integration
extends for $t = 2000$. The number of grid points is $256
\times 256$,
and the grid extends radially from $r_{\rm in} = 1.40$ to $r_{\rm
out}=300$ while it extends in the zenith angle from $\theta = 0$ to
$\theta = \pi$.

\begin{figure}
\epsscale{1.20}
\plotone{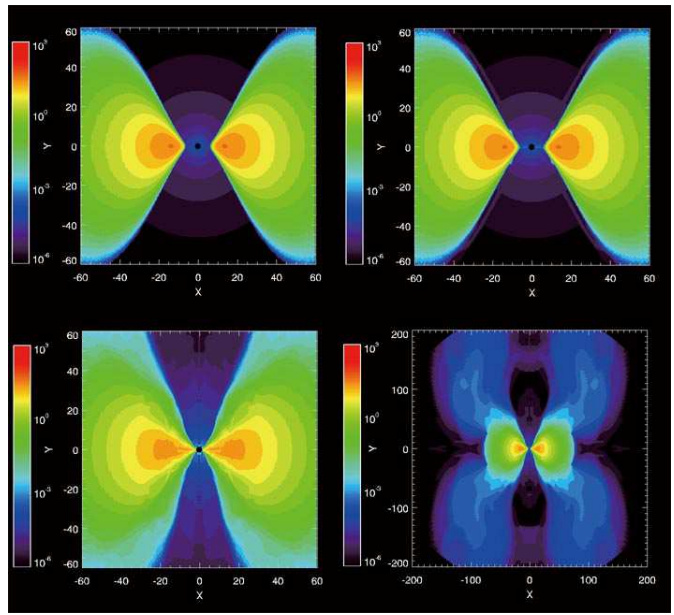}
\caption{Density contour in logarithmic scale (from $10^{-6}$ to
$10^3$) for the Fishbone-Moncrief
problem for the Kerr parameter $a=0.9$.
Number of grid points is $256 \times 256$. The simulations are done
until $t=2000$. The upper left panel shows the initial state. The
upper right panel shows the final state of the simulation without magnetic
fields. The lower left panel shows the result with magnetic
fields. The lower right panel is same with the lower left one, but for
a wide region. These results are projected on the (r sin
$\theta$,r cos $\theta$)-plane. 
\label{fig10}}
\end{figure}

The density contours in logarithmic scale (from $10^{-6}$ to
$10^3$) are shown in Fig.\ref{fig10}. These are projected on the (r
sin $\theta$,r cos $\theta$)-plane.
The upper left panel shows the initial state.
The upper right panel shows the final state of the simulation without magnetic
fields.
The lower left panel shows the final state of the simulation with magnetic
fields. The lower right panel is same with the lower left one, but for a
wide region. 
Due to the presence of the magnetic fields, the
angular momentum in the torus is conveyed outward and the the torus starts to
accrete, and the jet is launched from the BH around the polar region.
This result is consistent with the previous
studies~\citep{gammie03,mckinney04,mckinney06a,mckinney06}.  

\section{Simulation of Collapsar}\label{GRB1}

Since our code has passed the many test calculations shown in the previous
section, we now simulate the dynamics of a collapsar using the code.
However, we have to say beforehand that no microphysics is included in
the code such as nuclear reactions, neutrino processes, and equation of
state for dense matter. So this is the FIRST STEP of our project to
simulate the dynamics of a collapsar and formation of a relativistic
jet of a GRB. 

\begin{figure}
\epsscale{1.20}
\plotone{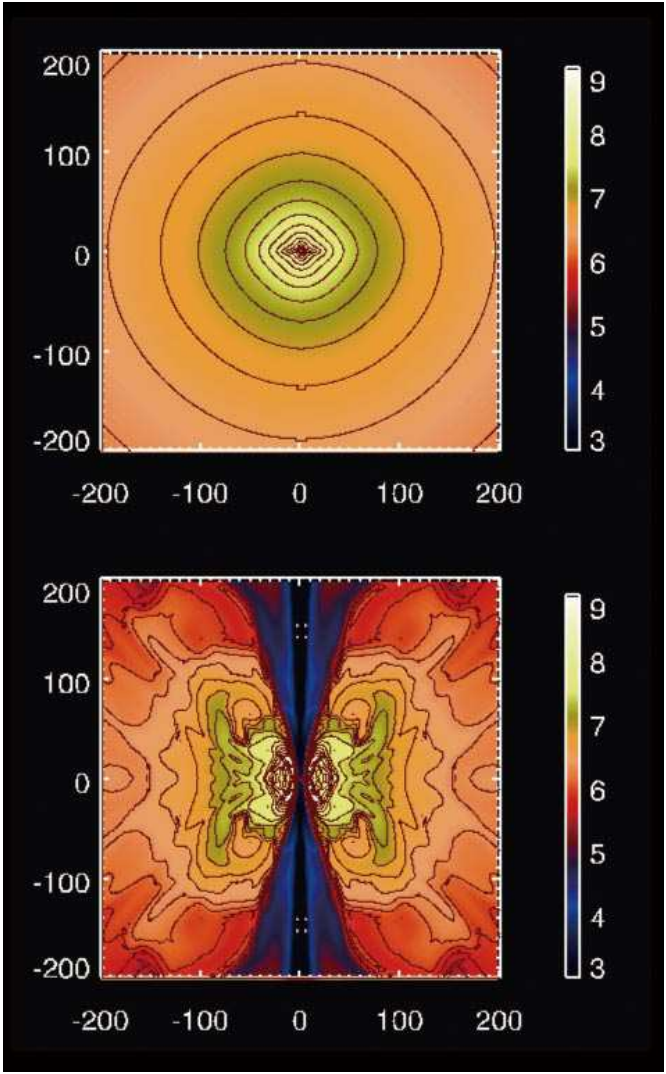}
\caption{Contours of rest mass density at the central region in
logarithmic scale, in which cgs units are used assuming that the
gravitational mass of the BH is 2$M_{\odot}$.
The length unit in the vertical/horizontal axes
corresponds to 2.95 $\times 10^5$ cm.
Upper panel (a) shows the state at $t=110000$ (that corresponds
to 1.0835 sec), while lower panel (b) shows the one at $t=180000$ (that
corresponds to 1.773 sec).
These results are projected on the (r sin
$\theta$,r cos $\theta$)-plane. 
\label{fig11}}
\end{figure}

\begin{figure}
\epsscale{1.20}
\plotone{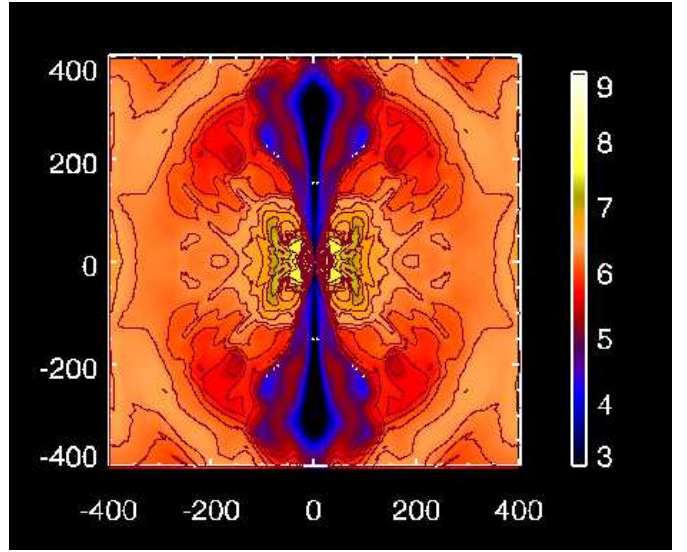}
\caption{Same with Fig.11(b), but for a wider region. 
\label{fig12}}
\end{figure}

\subsection{Method of Calculation}\label{GRB2}

We have done a 2D GRMHD simulation of a collapsar using
the Modified Kerr-Schild coordinate and $G=c=M=1$ units. 
When we show results, the coordinate is transfered from the
Modified Kerr-Schild coordinate to the Kerr-Schild one,
and the units are frequently transfered to cgs units.
The calculated region corresponds to a quarter of the
meridian plane under the assumption of axisymmetry and equatorial
symmetry. The spherical mesh with 256($r$)$\times$ 128($\theta$) grid
points is used for all the computations.
The calculated region covers from $r=$1.8 to 3$\times
10^{4}$ (that corresponds to 5.3$\times 10^{5}$cm and 8.9$\times
10^{9}$cm in cgs units, as explained below) with uniform grids in the
Modified Kerr-Schild space.

We adopt the model 12TJ in Woosley and Heger (2006).
This model corresponds to a star that has 12$M_{\odot}$ initially with
1$\%$ of solar metallicity, and rotates rapidly and does not lose its
angular momentum so much by adopting small mass loss rate. As a result, 
this star has a relatively large iron core of $1.82M_{\odot}$, and
rotates rapidly (the estimated Kerr parameter that a BH
forming of mass and angular momentum of the inner 3 $M_{\odot}$ would
formally have is 0.57) at the final stage. Of course, what kind of
stars are appropriate for progenitors of GRBs is still under
debate~\cite{yoon06}. Thus we chosed the model 12TJ as a first
example of our study because the iron core is large and rotating
rapidly, which seems to form a rapidly-rotating BH, among 
the models listed in Woosley and Heger (2006).
We assume that the central part of the star with 2$M_{\odot}$ has
collapsed and formed a BH at the center with the
Kerr parameter $a=0.5$. We also assume that the gravitational mass of
the BH is unchanged throughout the calculation.
Since $M=2M_{\odot}$, $r=1$ corresponds to
2.95 $\times 10^5$cm, as explained above. Also, the inner boundary
$r=1.8$ is set within the outer horizon $r_+ = 1 + \sqrt{1-a^2} = 1.866$.

Since 1-D calculation is done for the model 12TJ, we can use the data
directly only for the physical quanta on the equatorial plane. As for the
density, internal energy density, and radial velocity, we assume the
structure of the star is spherically symmetric. We also set
$u^\theta=0$ initially. As for $u^\phi$, we extrapolate its value such as
$u^\phi (r,\theta) = u^\phi (r, \pi/2) \times \sin \theta$.

Effects of magnetic fields are taken into account in the model 12TJ. 
However, again, since 1-D calculation is done, we do not know the
configuration of the magnetic fields. It is difficult to extrapolate
magnetic fields that satisfy the condition div$\bf B$ = 0 everywhere.
Also, there are much uncertainty on the amplitude of the magnetic
fields in a progenitor. Thus we do not use the information on magnetic
fields of the model 12TJ. Rather, we adopt the same treatment in
section \ref{test7}. That is, the vector potential $A_\phi \propto
{\rm max} (\rho/\rho_{\rm max} - 0.2,0) \sin^4 \theta  $ where
$\rho_{\rm max}$ is the
peak density in the progenitor (after extracting the central part of
the progenitor that has collapsed and formed a BH). The field is
normalized so that the minimum value of $p_{\rm gas}/p_{\rm mag}$ becomes
$10^2$. The definition of $p_{\rm mag}$ is
$p_{\rm mag} = b^2/2$.
The reason why we adopt the strong dependence on the zenith angle for
$A_\phi$ is so as not to suffer from discontinuity of magnetic fields
at the polar axis. The resulting biggest amplitude of the magnetic
fields is 7.4$\times 10^8$G at $r=950$ (2.8$\times 10^8$cm). 

We use a simple equation of state $p_{\rm gas} = (\gamma-1)u$ where we
set $\gamma$=4/3 so that the equation of state roughly represents
radiation gas. 

As for the boundary condition in the radial direction,
we adopt the outflow boundary condition for the inner and outer
boundaries~\cite{gammie03}. 
As for the boundary condition in the zenith angle direction, axis of
symmetry condition is adopted for the rotation axis, while
the reflecting boundary condition is adopted for the equatorial plane.
As for the magnetic fields, the equatorial symmetry boundary condition,
in which the normal component is continuous and the tangential component 
is reflected, is adopted.

\subsection{Results}\label{GRB3}

\begin{figure}
\epsscale{1.20}
\plotone{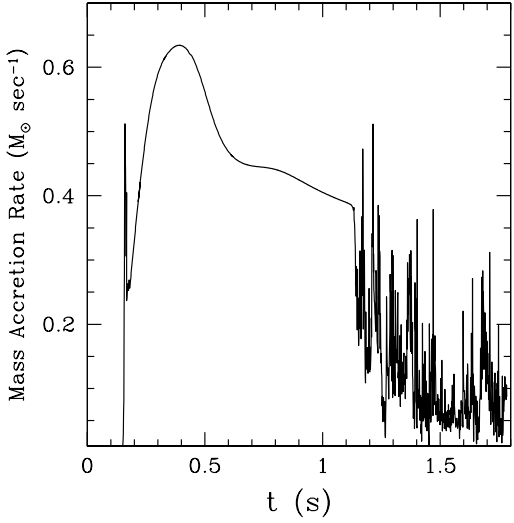}
\caption{Mass accretion rate history on the horizon. The unit
$M_{\odot}$ sec$^{-1}$ is used assuming that the
gravitational mass of the BH is 2$M_{\odot}$ throughout the
calculation.
\label{fig13}}
\end{figure}

In Fig.\ref{fig11}, color contours of rest mass density at the
central region are shown. Colors represent the density in units of
g cm$^{-3}$ in logarithmic scale. 
These results are projected on the (r sin$\theta$,r cos
$\theta$)-plane.
The length $r=200$ corresponds to 5.9 $\times 10^7$ cm.
The time unit corresponds to 9.85$\times 10^{-6}$ sec.
Upper
panel (a) represents the contours of rest mass density at $t=110000$ (that corresponds
to 1.0835 sec), while
lower panel shows the contours at $t=180000$ (that
corresponds to 1.773 sec).  
Fig.\ref{fig12} is the same figure with
Fig.\ref{fig11}(b), but for a wider region. A jet is clearly seen along
the rotation axis. In Fig.\ref{fig13}, mass accretion rate history on
the horizon is shown. The definition of the mass accretion rate is
\begin{equation}
\dot{M} = 2 \times  2\pi \int_{0}^{\theta} d \theta \sqrt{-g} \rho u^r.
\label{GRB3-0}
\end{equation}
It takes about 0.15 sec for the inner edge of
the matter to reach the horizon. When the matter reaches there, there
is an initial spike of the mass accretion rate. After that, there is a
quasi-steady state like Fig.\ref{fig11}(a) is realized. Then, the jet
is launched at $\sim$1.1 sec. After that, the mass accretion rate varies rapidly
with time, which resembles to a typical time profile of a GRB. 

\begin{figure}
\epsscale{1.20}
\plotone{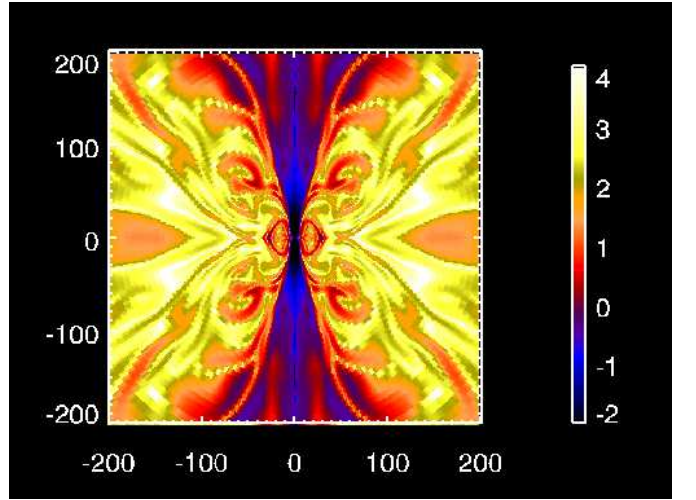}
\caption{Contour of the plasma beta ($p_{\rm gas}/p_{\rm mag}$) at
$t=180000$ in logarithmic scale.  
\label{fig14}}
\end{figure}

\begin{figure}
\epsscale{1.20}
\plotone{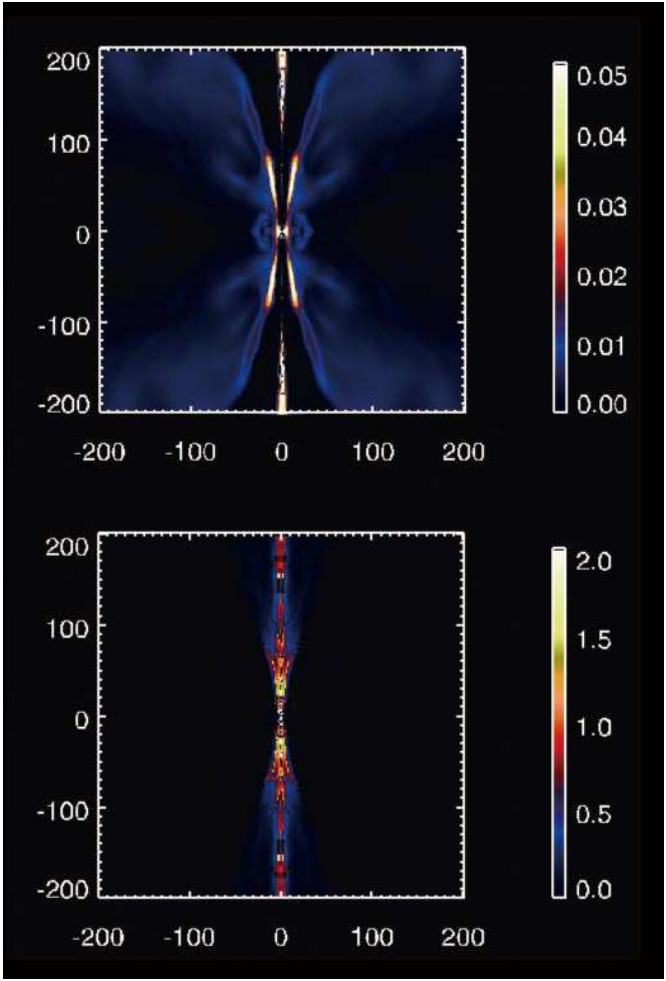}
\caption{Upper panel (a): contours of bulk Lorentz factor around the central region
at $t=180000$. Lower panel (b): contours of the energy flux per
unit rest-mass flux at $t=180000$ that represent the bulk Lorentz factor ($\Gamma_{\infty}$) of the invischid
fluid element when all of the internal and magnetic energy are
converted into kinetic energy at large distances. The contours are written in logarithmic scale.
\label{fig15}}
\end{figure}

We show color contours of the plasma beta ($p_{\rm gas}/p_{\rm mag}$)
in logarithmic scale at
$t=180000$ in Fig.\ref{fig14}. As expected, the plasma beta is low in the jet
region while it is high in the accretion disk region. 
We show color contours of bulk Lorentz factor around the central region
at $t=180000$ in Fig.\ref{fig15}(a) (upper panel, in logarithmic
scale). Color contours of the energy flux per
unit rest-mass flux ($E=-T^r_t /(\rho u^r)$), which is conserved for an inviscid fluid flow of
magnetized plasma, are also shown in Fig.\ref{fig15}(b) (lower panel, in logarithmic scale).
This value represents the bulk Lorentz factor ($\Gamma_{\infty}$) of the invischid
fluid element when all of the internal and magnetic energy are
converted into kinetic energy at large distances~\cite{mckinney06a}. We can see that the bulk Lorentz factor of the jet is
still low (Fig.\ref{fig15}(a)), but
it can be potentially as high as $10^2$ at large radius (Fig.\ref{fig15}(b)).
At $t=180000$, the strength of the magnetic field ($\sqrt{4 \pi b^2}$) at the
bottom of the jet is 
found to be $\sim 10^{15}$G, and $u^{\phi}$/$u^{t}$ is $\sim 0.1$ at $r_{\rm
ms}$ on the equatorial plane. Here $r_{\rm ms}$ is the 
marginally stable orbit. For the Kerr BH with $a=0.5$, $r_{\rm
ms}$ is 4.23.
As stated in section~\ref{GRB2}, the initial biggest amplitude of the
magnetic fields is
7.4$\times 10^8$G at $r=950$ where the initial density is $\sim 10^6$
g cm$^{-3}$, the expected amplification factor of the
magnetic fields due to the
gravitational collapse and differential rotation around the BH
is $(\rho / \rho_0)^{2/3} \times (d\Phi/dt / 2 \pi ) * \Delta t$ $\sim$ 100
$\times$ 0.016 $\times$ 180000 $\sim 3 \times 10^5$. Thus the initial magnetic field
can be amplified as large as several times of $10^{14}$ G, which is roughly consistent
with the amplitude of the magnetic fields at the bottom of the jet. At
late phase, the magneto-rotational instability (MRI) may be also
working, which is discussed in the next section.

\begin{figure}
\epsscale{1.20}
\plotone{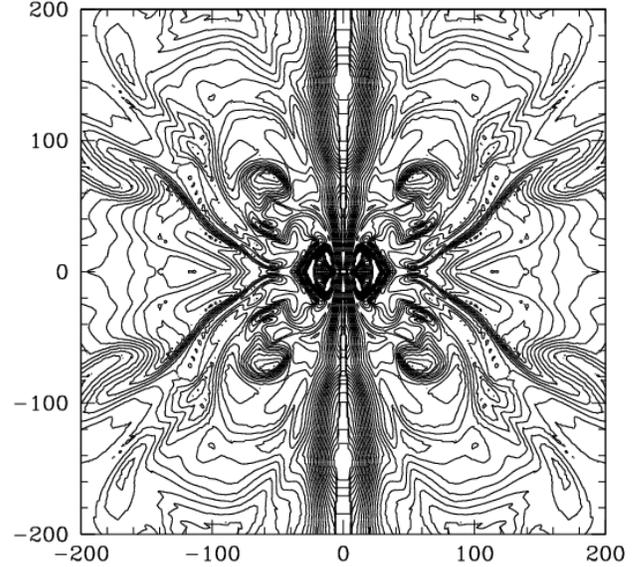}
\caption{Contours of the $\phi$ component of the vector potential
($A_\phi$) at $t=180000$. Level surfaces coincide with poloidal magnetic field
lines, and field line density corresponds to poloidal field
strength. The biggest amplitude of the magnetic fields is $\sim 10^{15}$G.   
\label{fig16}}
\end{figure}

In Fig.\ref{fig16}, contours of the $\phi$ component of the vector potential
($A_\phi$) at $t=180000$ are shown. Level surfaces coincide with
poloidal magnetic field lines, and field line density corresponds to
poloidal field strength. As expected, the magnetic fields are strong
at the jet region, which makes the plasma beta very low.
From Fig.\ref{fig16}, the opening angle of the jet is estimated as
$5^{\circ}-6^{\circ}$. From Fig.\ref{fig14}, \ref{fig15}(b), and
\ref{fig16}, this jet should correspond to the poynting flux
jet~\cite{hawley06}. This jet is surrounded by the funnel-wall jet
region~\cite{hawley06}, which is shown in Fig.\ref{fig17} below.

In Table.\ref{tab3} and \ref{tab4}, the integrated energies of matter and
electromagnetic field at $t=180000$ are shown. The
integrated region is between the horizon and
$r=200$ (for Table.\ref{tab3}) or $r=40$ (for Table.\ref{tab4}), and
within the zenith angle measured from the polar axis. 
As for the matter component, the contribution of the rest mass energy is
subtracted. That is,
\begin{equation}
E_{\rm Matter} = 2 \times  2\pi \int_{r_+}^{r = 200 \; \rm or \; 40} dr
\int_{0}^{\theta} d \theta \sqrt{-g} 
(T^0_{0, \rm Matter} - \rho u^0 u_0 ).
\label{GRB3-1}
\end{equation}
Factor 2 is coming from the symmetry of the system with respect to the
equatorial
plane. The field part can be written as
\begin{equation}
E_{\rm EM} = 4 \pi \int_{r_+}^{r = 200 \; \rm or \; 40} dr
\int_{0}^{\theta} d \theta \sqrt{-g} T^0_{0, \rm EM}.
\label{GRB3-2}
\end{equation}
It can be seen that the energy in electromagnetic field dominates that
in matter within $r \le 40$, while they become comparable within $r
\le 200$. Also, the integrated energy is still less than the typical
explosion energy of a GRB ($\sim 5 \times 10^{50}$erg; Frail et al. 2001).

Finally, we show the rest-mass density, outgoing mass flux, and
outgoing poynting flux in Fig.\ref{fig17}.
The top panel (a) shows the rest-mass density (g cm$^{-3}$) as a function of the zenith
angle at $r=10r_{\rm ms}=42.3$ and $t=180000$. It is seen that the low-density region is realized
around the polar axis, which corresponds to the jet region (0.1 radian
corresponds to 5.7$^{\circ}$). 
The middle panel (b) shows the outgoing mass flux $\rho
u^r$ (g cm$^{-2}$ s$^{-1}$) at $r=10r_{\rm ms}$ and $t=180000$. It is seen that the outgoing
mass flux exists around $0.06 \le \theta \le 0.14$, which
corresponds to the funnel-wall
jet~\citep{devilliers03b,hirose04,mckinney04,kato04,devilliers05,hawley06}.
The bottom panel (c) shows the outgoing poynting flux (in units of
10$^{50}$erg s$^{-1}$ rad$^{-1}$) at $r=r_+$ and $t=180000$. 
Definition of the outgoing poynting flux is
\begin{equation}
F_{\rm BZ} = - 2 \times 2 \pi \sqrt{-g} T^t_{r, \rm EM}.
\label{GRB3-3}
\end{equation}
Factor 2 is coming from the
assumption of the symmetry of the system with respect to the
equatorial plane. We can see that the positive outgoing poynting flux
exists at the jet region ($0 \le \theta \le 0.23$ in radian). 
The integrated energy of the outgoing poynting flux at $r=r_+$ and
$t=180000$ is $4.6 \times 10^{46}$ erg s$^{-1}$. Since the duration of the jet
in this study is $\sim 0.7$ s, this outgoing poynting flux seems to be
too weak to explain the energy of the jet listed in Table.\ref{tab3}
and \ref{tab4}. Thus we conclude that the jet is launched mainly by
the magnetic field amplified by the gravitational collapse and
differential rotation around the BH, rather than the
Blandford-Znajek mechanism in this study. 

\section{Summary and Discussion}\label{summary}

In order to investigate the formation of relativistic jets at the
center of a progenitor of a GRB, we have developed a two-dimensional
GRMHD code. In order to confirm the reliability of the code, we have
shown that the code passes many, well-known test calculations. Then 
we have performed a numerical simulation of a collapsar using a
realistic progenitor model.
We have followed the time evolution of the system for 1.773 s, and 
it was shown that a jet is launched from the center of the progenitor.
We also found that the mass accretion rate is in quasi-stable state
before the launch of the jet, while it shows rapid time variability
that resembles to a typical time profile of a GRB after the launch.
The structure of the jet is similar to the previous study: a poynting
flux jet is surrounded by a funnel-wall jet.
Even at the final stage of the simulation, the bulk Lorentz factor of the jet is still low, and the
total energy of the jet is still as small as $10^{48}$ erg. 
However, we found that the energy flux per
unit rest-mass flux ($E=-T^r_t /(\rho u^r)$) is as high as $10^2$ at the
at the bottom of the jet. 
Thus we conclude that the bulk Lorentz factor of the jet can be
potentially high when it propagates outward. Also, as long as the
duration of the activity of the central engine is long enough, the
total energy of the jet can be large enough to explain the typical
explosion energy of a GRB ($\sim 5 \times 10^{50}$ erg). It is shown that the
outgoing poynting flux exists at the horizon around the polar region,
which proves that the Blandford-Znajek mechanism is working. 
However, we conclude that the jet is launched mainly by
the magnetic field amplified by the gravitational collapse and
differential rotation around the BH, rather than the
Blandford-Znajek mechanism in this study.

\begin{figure}
\epsscale{1.20}
\plotone{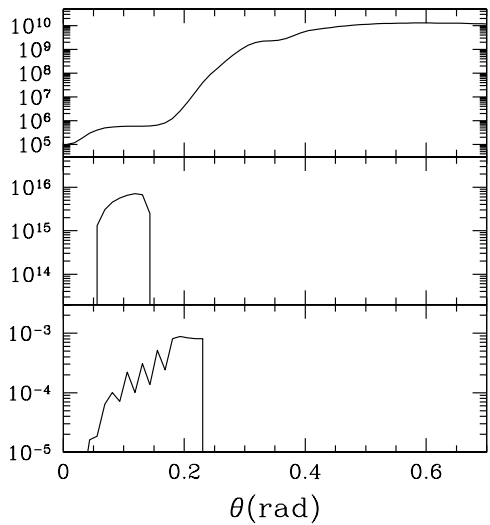}
\caption{
The top panel (a): the rest-mass density (g cm$^{-3}$) as a function of the zenith
angle at $r=10r_{\rm ms}=42.3$.
The middle panel (b): the outgoing mass flux $\rho
u^r$ (g cm$^{-2}$ s$^{-1}$) at $r=10r_{\rm ms}$. 
The bottom panel (c): the outgoing poynting flux (in units of
10$^{50}$erg s$^{-1}$ rad$^{-1}$) at $r=r_+$. These values are
evaluated at the final stage of the simulation ($t=180000$).  
\label{fig17}}
\end{figure}

When we apply the Blandford-Znajek formula~\cite{barkov08b}, the
integrated outgoing poynting flux is
\begin{eqnarray}
\dot{E} = 3.6 \times 10^{50} f(a) \Psi^2_{27} M_2^{-2} \;\; \rm erg
\; s^{-1}
\label{summary1}
\end{eqnarray}
where $M_{2} = M_{\rm BH}/2M_{\odot} = 1$, $\Psi_{27} = \Psi/10^{27}$
G cm$^2$, and $f(a) = a^2/(1 + \sqrt{1-a^2})^2 = 0.07179$.
This value becomes $2.3 \times 10^{46} B_{15}^2$ erg
s$^{-1}$ for the jet with opening angle $\theta = 5^{\circ}$, which is
comparable to our numerical result (Fig.\ref{fig17}(c)). 

As for the efficiency of converting the released gravitational energy
to the jet's energy, it can be estimated as follows: the mass
accretion rate is $\sim 0.1 M_{\odot}$ s$^{-1}$ (Fig.\ref{fig13}), the total energy of the
jet at the final stage is $\sim 10^{48}$ erg (Table.\ref{tab3}), and
the duration of the jet is $\sim 0.7$ s (Fig.\ref{fig13}). Thus the
efficiency can be estimated as $\sim 10^{-5}$. When we use
the outgoing poynting flux at the horizon (4.6$\times 10^{46}$erg s$^{-1}$),
the efficiency is as low as $\sim 3 \times 10^{-7}$. These values seem
to be very small compared with the previous
study~\citep{devilliers05,mckinney07a}. One of the reason will be
because they used an almost steady disk model. On the other hand, we
used a realistic progenitor model that collapses gravitationally. Thus
the resulting
mass accretion rate is pretty high. Second reason may be because the
efficiency is still low even at the final stage of the simulation.
If we perform the simulation further,
the efficiency may become higher with time: mass accretion rate will become
smaller, and the jet energy might be larger due to the amplification
of the magnetic fields due to winding-up (and MRI) effects. Also, when
the initial amplitude of the magnetic field is set to be larger (as in
Barkov and Komissarov 2008), the efficiency may be enhanced. 
Further, we should investigate the dependence of the dynamics on
progenitor models as well as the Kerr parameter of the BH.
We are planning to investigate this point
systematically in the next paper.

It is well known that the system is unstable against
MRI when there is a strong negative
shear profile $(d \Omega/d \ln r)$~\citep{balbus91,balbus94}, where
$\Omega$ is the angular velocity. 
The saturation toroidal magnetic field strength is roughly expected to
be $B_{\phi} \sim (4 \pi \rho)^{1/2} r
\Omega$~\citep{akiyama03,akiyama05}, which is confirmed by semi-global
simulations~\cite{obergaulinger08}. The saturation poloidal magnetic
field strength is roughly an order of magnitude
smaller~\cite{obergaulinger06}. Thus $B_{\phi}$ may be amplified by
MRI as strong as $1.4 \times 10^{15}$G $\rho_9 r_{\rm ms}
\Omega_{4}$ where $\Omega_4$ is $\Omega / 10^4$ rad s$^{-1}$.
The characteristic
timescale for saturating the MRI is the Alfv$\rm \acute{e}$n crossing
time: $t_A \sim 0.1$ms $R_6 \rho_9^{1/2} B_{15}^{-1}$ where $R_6$,
$\rho_9$, and $B_{15}$ are the
radius in units of $10^6$cm, the density in units of $10^9$ g cm$^{-3}$, and
the amplitude of magnetic fields in units of $10^{15}$G, respectively.
Thus this characteristic timescale can be
shorter than the winding-up timescale for strong magnetic fields. 
However, the length scale of the mode with the largest MRI growth rate
is approximately $\lambda_{\rm MRI} \sim 50$ cm $P_{0.5}
B_{10} \rho_9^{-1/2}$ where $P_{0.5}$ is the rotation period in units
of 0.5 ms, which is too short to be resolved numerically. At least, it
is not resolved in the beginning of the simulations. After the
magnetic field is amplified to a certain value due to gravitational
collapse and winding-up effect, MRI may be working in this
study~\citep{obergaulinger06b,ott06,burrows07,dessart08}.  
It will be necessary to develop a sophisticated code that takes into
account the MRI effectively with help of semi-global
simulations~\cite{obergaulinger08} in order to evaluate the
influence of MRI on the dynamics of a collapsar.

It is well known that it becomes difficult to obtain
the matter part of the primitive variables ($\rho$,$u$,$u^{i}$)
precisely by the Newton-Raphson method~\cite{noble06}
due to numerical truncation
errors~\citep{komissarov02,komissarov04a,komissarov04b,mckinney04,komissarov05,mckinney06}
when the electromagnetic part of the stress energy tensor (
$T^{\mu \nu}_{\rm EM}$) greatly exceeds the matter part (
$T^{\mu \nu}_{\rm Matter}$).
The problem is that the time integration of the electromagnetic part 
does not become so reliable, either. This is because the MHD condition
($u_\mu F^{\mu \nu}=0$) is implicitly assumed in the basic equations,
and the resulting basic equation of electromagnetic part depends on
the velocity of fluid (Eq.\ref{eq3}).
Such pathological conditions may be realized at the bottom of the jet in our
study. In order to confirm the validity of our results in this study,
we are planning to develop a general relativistic force-free code that
is coupled with the GRMHD code sophisticatedly~\cite{mckinney06}.

It is very important to evaluate the terminal bulk Lorentz factor,
because GRBs are considered to be emissions from relativistic flows
with their bulk Lorentz factors greater than
$10^2$~(e.g. \cite{lithwick01}). Although an ad-hoc thermal (and
kinetic) energy deposition into the polar region seems to lead to
relativistic jets with bulk Lorentz factors $\sim
100$~\citep{aloy00,aloy02,zhang03,zhang04,cannizzo04,mizuta06,mizuta08,morsony07,wang08},
it is still controversial whether such ad-hoc energy deposition is
justified by numerical simulations with proper neutrino
physics~\cite{nagataki07}. On the other hand, numerical study on the
acceleration of electromagnetically powered jet requires quite
high-resolution~\citep{komissarov07b,narayan07,tchekhovskoy08,komissarov08}.
Due to the reason, a simplified jet model with an idealized boundary
condition is used at present in order to investigate whether the
initial poynting flux can be effectively converted into kinetic
energy~\citep{tchekhovskoy08,komissarov08}. According to their
results, as long as confinement of the jet is realized, acceleration
operates over several decades in radius and considerable
fraction of the poynting flux can be converted into the kinetic
energy. Thus, from the high-ratio of the poynting flux relative to
rest-mass flux seen in our study (Fig.\ref{fig15}(b)), a
relativistic jet with high bulk Lorentz factor may be realized at
large radius.

It is true that the two-dimensional restriction can be a significant
limitation. First, anti-dynamo theorem~\cite{moffat78} prevents the
indefinite maintenance of the poloidal magnetic field in the face of
dissipation. Second, axisymmetric simulations tend to overemphasize
the channel mode~\cite{hawley92}, which produces coherent internal
magnetized flows rather than the more generic MHD
turbulence. Hydrodynamic instability in the azimuthal direction may be
also very important~\citep{nagakura08,nagakura09}. Thus we
are planning to develop a three-dimensional GRMHD code~\citep{devilliers03b,hirose04,devilliers05,hawley06,beckwith08,shafee08,mckinney08} and investigate
the difference between two-dimensional simulations of collapsars
with three-dimensional ones.

In this study, we assumed that the central region of the progenitor
has collapsed and a BH is formed at the center with
surrounding envelope unchanged. Thus we solved the GRMHD equation on a
fixed background. But the final goal of our project is
to study how a GRB is formed from the gravitational collapse of a
massive star. Thus we are planning to develop a GRMHD code on a
dynamical background, which makes the study on the gravitational
collapse and BH formation at the center of a massive star
possible~\citep{shibata03,sekiguchi04,sekiguchi05,baiotti05,duez06,sekiguchi07}.

In this study, photo-disintegration of nuclei and neutrino processes
are not taken into account. Photo-disintegration absorb considerable
amount of thermal energy, and cooling/heating due to neutrino
processes may have great influence on the dynamics of a collapsar~\citep{dimatteo02,kohri02,nagataki03a,surman04,lee05,gu06,nagataki07,kawanaka07,kawabata08,rossi08,zhang09,cannizzo09}.
Especially, pair-annihilation of electron-type neutrinos may be a key
process to drive a GRB jet~\citep{woosley93,macfadyen99,asano00,asano01,miller03,surman05,kneller06,shibata07,birkl07}. Thus we are
planning to include such microphysics in our code, and perform more
realistic simulations of collapsars.

The SNe associated with GRBs often show peculiar properties.
Some are very energetic and
blight~\citep{galama98,iwamoto98,hjorth03,malesani04},
but others prohibit such blight
SNe from being accompanied~\citep{fynbo06,dellavalle06,galyam06}.
Since the brightness of SNe depends on
the mass of produced $\rm ^{56}Ni$~\citep{woosley99,nakamura01}, it is suggested that there is
a huge variety of the amount of $\rm ^{56}Ni$ in a SN
that associates with a GRB. It is still controversial
where and when $\rm ^{56}Ni$ is produced in a SN accompanied by
a GRB~\citep{nagataki03,nagataki06}.
It may be produced in a GRB
jet~\citep{maeda02,maeda03,tanaka07,maeda08,tominaga09,maeda09,bucciantini09}, or it
may be produced in (or outflow from) the
accretion disk around the BH~\citep{macfadyen99,pruet04,fujimoto04,surman06,hu08}, or it may
be synthesized around a proto-neutron star~\cite{uzdensky07}. At present, it is impossible to
investigate the explosive nucleosynthesis in a collapsar in our code
because nuclear reactions are not taken into account. We are planning
to include this effect, and study the site of $\rm ^{56}Ni$ production.
Also, study of a GRB as a possible site where very heavy elements and
light elements are synthesized is very important~\citep{lemoine02,beloborodov03,suzuki05}.

\acknowledgments
This research was supported by Grant-in-Aid for Scientific Research on
Priority Areas No. 19047004 by Ministry of Education, Culture, Sports,
Science and Technology (MEXT), Grant-in-Aid for Scientific Research (S)
No. 19104006 by Japan Society for the Promotion of Science (JSPS), and
Grant-in-Aid for young Scientists (B) No.19740139 by Japan Society for
the Promotion of Science (JSPS). The computation was carried
out on NEC SX-8 at Yukawa Institute for Theoretical Physics (YITP) in
Kyoto University and Cray XT4 at Center for Computational Astrophysics
(CfCA) in National Astronomical Observatory of Japan (NAOJ).
A portion of this work was carried out while the author was at the
Kavli Institute for Particle Astrophysics and Cosmology (KIPAC).
I am grateful to Roger Blandford, the director of KIPAC, for continuous
encouragement, useful discussions, and warm hospitality at KIPAC.
I also thank Weiqun Zhang,
Jonathan C. McKinney, Dmitri A. Uzdensky, Ruben Krasnopolsky for
useful discussions. I wish to thank all of the colleagues in YITP and
KIPAC for useful discussions and kind support.   
Finally, I would like to thank my family for continuous, warm support.

\begin{table}
\begin{center}
\caption{Initial Conditions for Shock Tube Test1 and Collision Test
\label{tab1}}  
\begin{tabular}{clrrrrrrrrrr}
\tableline\tableline
Test Type &  & $\rho$ & $p$ & $v^x$ & $v^y$ & $v^z$ & $B^x$ &
$B^y$ & $B^z$  \\
\tableline
Shock Tube Test1 & left state  & 1 & 1000  & 0 & 0 & 0  & 1 & 0 & 0 \\
                 & right state & 0.1 & 1   & 0 & 0 & 0  & 1 & 0 & 0 \\
\tableline
Collision Test   & left state  & 1   & 1   &  5/$\sqrt{26}$ & 0  & 0  &
                 10 & 10 & 0   \\
                 & right state & 1   & 1   & -5/$\sqrt{26}$ & 0  & 0
                 & 10 & -10 & 0 \\
\tableline
\end{tabular}
\tablecomments{$\gamma$ for the equation of state is set to be
                 4/3. Final time is set to be 1.0 for Shock Tube
                 Test 1, and 1.2 for Collision Test.}
\end{center}
\end{table}

\begin{table}
\begin{center}
\caption{Initial Conditions for 2D Shock Tube Problem
\label{tab2}}  
\begin{tabular}{cccrrrrrrrrr}
\tableline\tableline
Region & $x$  & $y$  & $\rho$ & $p$ & $v^x$ & $v^y$   \\
\tableline
A & 0 $\leqq$ $x$ $\leqq$ 0.5  & 0.5 $\leqq$ $y$ $\leqq$ 1  & 0.1  & 1 & 0.99 & 0  \\
B & 0.5 $\leqq$ $x$ $\leqq$ 1  & 0.5 $\leqq$ $y$ $\leqq$ 1  & 0.1  & 0.01 & 0 & 0  \\
C & 0 $\leqq$ $x$ $\leqq$ 0.5  & 0 $\leqq$ $y$ $\leqq$ 0.5  & 0.5  & 1 & 0 & 0  \\
D & 0.5 $\leqq$ $x$ $\leqq$ 1  & 0 $\leqq$ $y$ $\leqq$ 0.5  & 0.1  & 1 & 0 & 0.99  \\
\tableline
\end{tabular}
\tablecomments{$\gamma$ for the equation of state is set to be
                 5/3. Final time is set to be 0.4.}
\end{center}
\end{table}

\begin{table}
\begin{center}
\caption{Integrated Energies of Matter and Field (r$\le$200)
\label{tab3}}  
\begin{tabular}{cccrrrrrrrrr}
\tableline\tableline
$\theta$ & $0.714^{\circ}$  & $1.43^{\circ}$  & $2.14^{\circ}$ & $2.86^{\circ}$ & $3.57^{\circ}$   \\
\tableline
Matter   & 1.44E+46     & 7.09E+46      & 1.70E+47     & 3.14E+47     & 5.10E+47  \\
Field    & 2.96E+46     & 1.11E+47      & 2.39E+47     & 4.12E+47     & 6.30E+47  \\
\tableline\tableline
$\theta$ & $4.29^{\circ}$  & $5.00^{\circ}$  & $5.71^{\circ}$ & $6.42^{\circ}$ & $7.14^{\circ}$   \\
\tableline
Matter   & 7.66E+47      & 1.10E+48      & 1.52E+48     & 2.05E+48     & 2.69E+48   \\
Field    & 8.91E+47      & 1.19E+48      & 1.52E+48     & 1.88E+48     & 2.26E+48   \\
\tableline
\end{tabular}
\tablecomments{The energy is written in units of erg. As for the
matter component, the contribution of the rest mass energy is
subtracted.}
\end{center}
\end{table}

\begin{table}
\begin{center}
\caption{Integrated Energies of Matter and Field (r$\le$40)
\label{tab4}}  
\begin{tabular}{cccrrrrrrrrr}
\tableline\tableline
$\theta$ & $0.714^{\circ}$  & $1.43^{\circ}$  & $2.14^{\circ}$ & $2.86^{\circ}$ & $3.57^{\circ}$   \\
\tableline
Matter   & 6.89E+43     & 3.15E+44      & 8.96E+44     & 2.03E+45     & 3.95E+45  \\
Field    & 8.60E+45     & 3.44E+46      & 7.73E+46     & 1.37E+47     & 2.14E+47  \\
\tableline\tableline
$\theta$ & $4.29^{\circ}$  & $5.00^{\circ}$  & $5.71^{\circ}$ & $6.42^{\circ}$ & $7.14^{\circ}$   \\
\tableline
Matter   & 6.76E+45      & 1.04E+46      & 1.47E+46     & 1.96E+46     & 2.50E+46   \\
Field    & 3.08E+47      & 4.19E+47      & 5.46E+47     & 6.91E+47     & 8.52E+47   \\
\tableline
\end{tabular}
\tablecomments{Same with Table.3, but integration is done for R$\le$40.}
\end{center}
\end{table}

\end{document}